\documentclass[aps,prl, twocolumn,superscriptaddress]{revtex4-1}
\usepackage{amsfonts}
\usepackage{amsmath}
\usepackage{amssymb}
\usepackage{graphicx}
\usepackage{times}
\usepackage{physics}
\usepackage[colorlinks=true,linkcolor=blue,citecolor=red,plainpages=false,pdfpagelabels]
{hyperref}
\usepackage{color}

\providecommand{\U}[1]{\protect\rule{.1in}{.1in}}

\begin{document}

\title{What are Quantum Temporal Correlations?}
\author{Hai Wang}
 \affiliation{School of Mathematical Sciences, Zhejiang University, Hangzhou 310027, PR~China}
 \author{Ray-Kuang Lee}
 \email{rklee@ee.nthu.edu.tw}
\affiliation{Institute of Photonics Technologies, National Tsing Hua University, Hsinchu 30013, Taiwan \\
 Physics Division, National Center for Theoretical Sciences, Hsinchu 30013, Taiwan}
\author{Manish Kumar Shukla}
 \affiliation{Center for Computational Natural Sciences and Bioinformatics, \\
 International Institute of Information Technology, Gachibowli, Hyderabad, India}
\author{\\ Indranil Chakrabarty}
  \affiliation{Center for Security Theory and Algorithmic Research, \\
 International Institute of Information Technology, Gachibowli, Hyderabad, India}
 \author{Shao-Ming Fei}
 \affiliation{School of Mathematical Sciences, Capital Normal University, Beijing 100048, PR~China, \\
 Max-Planck-Institute for Mathematics in the Sciences, 04103 Leipzig, Germany}
 \author{Junde Wu}
 \email{wjd@zju.edu.cn}
 \affiliation{School of Mathematical Sciences, Zhejiang University, Hangzhou 310027, PR~China}

\date{\today}
\begin{abstract}
Space and time are crucial twins in physics. In quantum mechanics, spatial correlations already reveal nonclassical
features, such as entanglement, and have bred many quantum technologies. However,  the nature of quantum temporal correlations still remains in vague.
In this Letter, based on the entangled-history formalism, we  prove rigorously that temporal correlations are equivalent to spatial correlations.
The effect of temporal correlations corresponds to a quantum channel. The resulting quantifications and classifications of quantum temporal correlations are illustrated in a natural way.
Our proposed procedures also show  how to determine temporal correlations completely.
\end{abstract}

\maketitle

Quantum entanglement was firstly introduced by E. Schr\"odinger in 1935~\cite{epr, horo}, which describe the spatial correlations of particles. After the famous work of J. S. Bell~\cite{bell}, people started to appreciate the features of quantum spatial correlations, which  distinguish quantum mechanics from classic physics and provide the foundation for quantum technologies~\cite{horo1, Al, horo2, cs, rs}. On the other hand, a single particle  \textmd{W} different instants can also reveal nonclassical correlations, which are termed as quantum temporal correlations. In fact, as early as in 1985,  by performing measurements at four different instants, the Leggett-Garg inequality~\cite{LG} was established to show the temporal correlations~\cite{JH1,JH2, JH3, JH4, JH5}. Nevertheless,  due to the strong action of measurements, what is the physical essence of quantum temporal correlations  still remains open.

In order to answer what temporal correlations are, Y. Aharonov et al take the causal relation into the time-symmetric interpretations of quantum mechanics and proposed the ``multiple-time states" theory~\cite{yak, yak2, yak3, NCH}.
Alternatively,  J. Cotler and F. Wilczek established the entangled-history formalism~\cite{frank1,  frank3, frank4}.
In particular, based on the entangled-history formalism~\cite{frank2}, they showed that temporal CHSH inequality can be constructed in comparison with the famous spatial CHSH inequality~\cite{chsh}.
More recently, F. Costa et al. manifested that quantum spatial correlations and quantum temporal correlations can be transformed between each other~\cite{Costa18}.
Nevertheless, this transformation is still unable to tell us what quantum temporal correlations are.

As a discrete form to Feynman's path integral~\cite{path1},  the essence of  the entangled-history formalism is the creation of the entanglement structure of system evolution in time through the superposition of evolution paths.
Due to the causal nature of time evolution, how to have  {\it meaningful}  history operators is the most challengeable.
In this Letter, based on the entangled-history formalism, we show that quantum temporal correlations are spatial correlations in nature. Moreover, we  prove rigorously  that the effect of temporal correlations between two instants is a quantum channel.
With quantum state tomography, this quantum channel can be completely determined. By considering the entanglement property of this channel, we can quantify the quantum temporal correlations.

Let $t_{0}$ and $t_{1}$ be two different instants, $t_0 < t_1$. At $t_{0}$ and $t_{1}$, quantum systems of the particle \textmd{W} are represented by the Hilbert space $H_0$ and $H_1$, whose dimensions are $d_{0}$ and $d_{1}$, respectively. The effect of temporal correlation between $t_{0}$ and $t_{1}$ can be seen as a map $\Phi$ from $L(H_0)$ to $L(H_1)$, which maps quantum states in $H_{0}$ to quantum states in $H_{1}$.
By the superposition principle of quantum mechanics, we can assume that $\Phi$ is linear~\cite{linear}. Since $\Phi$ maps quantum states into quantum states, thus, it is easy to show that $\Phi$ is a trace-preserving positive linear map. In the following, we show that $\Phi$ is a completely positive trace-preserving linear map. That is, $\Phi$ is a quantum channel.

Firstly, we show that temporal correlations are just spatial correlations.

Let $\{\ket{\alpha_{i}}\}$ and $\{\ket{\beta_{j}}\}$ be orthonormal bases of $H_{0}$ and $H_{1}$, respectively. Then, $\{E_{ij}\equiv \ket{\alpha_{i}}\bra{\alpha_{j}}, \forall
i,j\}$ and $\{F_{kl}\equiv \ket{\beta_{k}}\bra{\beta_{l}}, \forall
k,l\}$ are bases of $L(H_0)$ and $L(H_1)$, respectively. If $\rho\in L(H_{0})$ is a quantum state on $H_0$, then $\rho=\sum_{i,j}\rho_{ij}E_{ij}$.
Now, we introduce two auxiliary systems $H_A$ and $H_B$ to the system $H_0$, where $H_A$ is used to store information of $H_0$, while $H_B$ to store information of $H_1$.
Here,  $dim(H_{A})=dim(H_{0})=d_{0}$, $dim(H_{B})=dim(H_{1})=d_{1}$. The auxiliary systems are initialized in pure states $\ket{0_{A}}$ and $\ket{0_{B}}$, respectively.
If $\rho$ is the initial state of $H_{0}$, then the initialized state of the composite system $H_0\otimes H_B\otimes H_A$ is
\begin{eqnarray}
\widetilde{\rho}_{0}\doteq\sum_{i,j}\rho_{ij}E_{ij}\otimes \ket{0_{B}}\bra{0_{B}}\otimes
\ket{0_{A}}\bra{0_{A}}.
\end{eqnarray}

Now, we illustrate the procedures to characterize the quantum correlations between two instants.
Firstly, we perform a unitary operation on $H_{0}\otimes H_A$ by
$U_{0}\ket{\alpha_{i}}\ket{0_{A}}=\ket{\alpha_{i}}\ket{\alpha_{i}},
\forall i$. Then at $t_{0}$, the new quantum state of $H_0\otimes H_B\otimes H_A$ is
\begin{eqnarray}
\widetilde{\rho}_{1}\doteq U_{0}\widetilde{\rho}_{0}U^{\dagger}_{0}=\sum_{i,j}\rho_{ij}E_{ij}\otimes \ket{0_{B}}\bra{0_{B}}\otimes E_{ij}.
\end{eqnarray}
At $t_{1}$, with the effect of temporal correlation $\Phi$,  one can have  the quantum state
\begin{eqnarray}
\widetilde{\rho}_{2} &\doteq&\sum_{i,j}\rho_{ij}\Phi(E_{ij})\otimes
\ket{0_{B}}\bra{0_{B}}\otimes E_{ij},\\
&=&\sum_{i,j,k,l}\rho_{ij}\Phi_{kl,ij}F_{kl}\otimes
\ket{0_{B}}\bra{0_{B}}\otimes E_{ij},
\end{eqnarray}
where $\Phi_{kl,ij}=\text{Tr}(F_{kl}^{\dagger}\Phi(E_{ij}))$. Now, we perform another unitary operation on $H_{1}\otimes H_B$ by
$U_{1}\ket{\beta_{k}}\ket{0_{B}}=\ket{\beta_{k}}\ket{\beta_{k}},
\forall k$. Then, the quantum state becomes
\begin{eqnarray}
\widetilde{\rho}_{3}\doteq U_{1}\widetilde{\rho}_{2}U_{1}^{\dagger}=\sum_{i,j,k,l}\rho_{ij}\Phi_{kl,ij}F_{kl}\otimes F_{kl}\otimes
E_{ij}.
\end{eqnarray}

If we choose $\ket{\gamma_{0}}=\frac{1}{\sqrt{d_{1}}}\sum_{j}\ket{\beta_{j}}$, then it can be shown that $\bra{\gamma_{0}}\widetilde{\rho}_{3}\ket{\gamma_{0}}=\frac{1}{d_{1}}$. Thus, we can perform a local quantum operation on the system $H_{1}$ via the Kraus element $\ket{\gamma_{0}}\bra{\gamma_{0}}$. Then the quantum state on the system $H_B\otimes H_A$  is reduced to
 \begin{equation}
\widetilde{\rho}=\sum_{i,j}\sum_{k,l}\rho_{ij}\Phi_{kl,ij}F_{kl}\otimes E_{ij}.
\end{equation}
Note that Eq. (6) is a bipartite state {\it in space}, which shows an important physical fact: the temporal correlation between $t_{0}$ and $t_{1}$ is stored in the spatial bipartite state $\widetilde{\rho}$.

As a consequence, the correlation of a particle W at two different instants is just the spatial correlation of two subsystems at two different places. That is, temporal correlations are equivalent to spatial correlations.

On the other hand, it is known that a positive trace-preserving linear map is a quantum channel if and only if it is completely positive.
In order to show that the effect of temporal correlation $\Phi$ is a quantum channel, we only need to prove that it is completely positive.
With Choi theorem, we know that $\Phi$ is completely positive if and only if the matrix $$\sum_{i,j,k,l}\Phi_{kl,ij}F_{kl}\otimes E_{ij}$$ is a positive matrix, coined as the Choi matrix~\cite{choi}.
On the other hand, the equation (6) shows that for each initial state $\rho$,  its
$\widetilde{\rho}$ is a quantum state. In particular, if we  take $\ket{\mu}= \frac{1}{\sqrt{d_{0}}}\sum_{i}\ket{\alpha_{i}}$ into Eq.(6), then we will get that\begin{equation}
\rho_{\mu, \Phi}\equiv\frac{1}{d_0}\sum_{i,j,k,l}\Phi_{kl,ij}F_{kl}\otimes E_{ij}
\end{equation} is a quantum state. Therefore, $$\sum_{i,j,k,l}\Phi_{kl,ij}F_{kl}\otimes E_{ij}$$ is a positive matrix. Thus, we proved that $\Phi$ is a quantum channel.

Moreover, if we use tomography technique~\cite{tomo} to determine the  quantum state $\rho_{\mu, \Phi}$, then we can decide the matrix $(\Phi_{kl,ij})_{kl,ij}$ completely. That is, we can decide $\Phi$ completely.
Henceforth, without losing generality,
we assume $d_{0}=d_{1}=d$.

\

\textit{Quantifying temporal correlations}:  Based on our main result given in Eq. (6), one can naturally apply the entanglement formation $E(\cdot)$ of the bipartite state $\rho_{\mu,\Phi}$ to quantify temporal correlations~\cite{formation1, formation2, formation3, formation4, continuity}.
For the given bases $\{\ket{\alpha_{i}}\}$ and $\{\ket{\beta_{j}}\}$ of $H_{0}$ and
  $H_{1}$, we define
\begin{eqnarray}
Q_{\{\ket{\alpha_{i}}\},\{\ket{\beta_{j}}\}}(\Phi)=E(\rho_{\mu, \Phi}),
\end{eqnarray}
as a figure of metric to measure the temporal correlation $\Phi$ with respect to $\{\ket{\alpha_{i}}\}$ and
$\{\ket{\beta_{j}}\}$.
To avoid the dependence on the choices of orthonormal bases, we set
\begin{eqnarray}
Q(\Phi)=
\inf_{\{\ket{\alpha_{i}}\},\{\ket{\beta_{j}}\}}
Q_{\{\ket{\alpha_{i}}\},\{\ket{\beta_{j}}\}}(\Phi),
\end{eqnarray}
where the $\inf$ runs over all possible orthonormal bases
$\{\ket{\alpha_{i}}\}$ and $\{\ket{\beta_{j}}\}$ of $H_{0}$ and $H_{1}$, respectively, in order to quantity the temporal correlation.
It follows from the definition of $Q(\Phi)$ given in Eq. (9) that its range is $[0, \log\,d]$.
Moreover, from the invariance of the entanglement of formation under the local unitary transformation, we have
\begin{equation}
Q(\Phi)=Q(\mathcal{V}\circ \Phi),
\end{equation}
where $\mathcal{V}$ is an arbitrary quantum unitary operation on $L(H_{1})$, and $\circ$ represents the composition of maps.
Based on the entanglement formation, in the following,  the strongest and the weakest temporal correlations are both studied as an illustration.

\

\textit{ The strongest temporal correlations:} First of all, we give an important conclusion: The quantum state $\rho_{\mu,\Phi}$ is a maximally entangled state on $H_{1}\otimes H_{0}$ if and only if $\Phi$, the effect of temporal correlations, is a quantum unitary channel.

To prove this statement, note that Eq. (7) can be re-written as
\begin{equation}
\rho_{\mu,\Phi}=\frac{1}{d}\sum_{i,j,k,l}\Phi_{kl,ij}F_{kl}\otimes E_{ij}=\Phi\otimes I(\frac{1}{d}\sum_{i,j}\ket{\alpha_{i}\alpha_{i}}\bra{\alpha_{j}\alpha_{j}}).
\end{equation}
Thus, if $\Phi$ is a quantum unitary channel, it is obvious that $\rho_{\mu,\Phi}$ is a maximally entangled state and the sufficiency is proved.

On the other hand, if $\rho_{\mu,\Phi}$ is a maximally entangled state, then we can find orthonormal bases $\{\alpha_{i}\}_{i}$ and $\{\beta_{j}\}_{j}$ on $H_{0}$ and $H_{1}$ such that
\begin{eqnarray*}
\rho_{\mu,\Phi}=\ket{\gamma}\bra{\gamma},
\end{eqnarray*}
where $\ket{\gamma}=\frac{1}{\sqrt{d}}\sum_{i}\ket{\beta_{i}\alpha_{i}}.$
Now, one can construct a unitary operator $U\in L(H_{0},H_{1})$ such that $U\ket{\alpha_{i}}=\ket{\beta_{i}},\forall i,$ and the unitary operator define a quantum unitary operation $\mathcal{U}$. It is easy to show that $\Phi$ and $\mathcal{U}$ have the same Choi matrix.
As a result from the Choi-Jamilkowski isomorphism theorem, we have $\Phi = \mathcal{U}$, giving that $\Phi$ is a quantum unitary channel. The necessity is proved.

Furthermore, for each unitary channel, it can be verified easily that for arbitrary orthonormal bases $\{\ket{\alpha_{i}}\}$ on $H_{0}$ and $\{\ket{\beta_{j}}\}$ on $H_{1}$, the quantum state $\rho_{\mu,\Phi}$ is always a maximally entangled state.
In particular, it means that for arbitrary orthonormal bases $\{\ket{\alpha_{i}}\}$ on $H_{0}$ and $\{\ket{\beta_{j}}\}$ on $H_{1}$, the quantity $Q_{\{\ket{\alpha_{i}}\},\{\ket{\beta_{j}}\}}(\Phi)$ is
always $\log d$. This showed that if $\Phi$ is a unitary channel, then $Q(\Phi)=\log d$.
On the other hand, Ref.~\cite{formation4} showed that for arbitrary orthonormal bases $\{\ket{\alpha_{i}}\}$ on $H_{0}$ and $\{\ket{\beta_{j}}\}$ on $H_{1}$, $Q_{\{\ket{\alpha_{i}}\},\{\ket{\beta_{j}}\}}(\Phi)=\log d$ means that the quantum state $\rho_{\mu,\Phi}$ is a maximally entangled state.
So,  $Q(\Phi)$ gets its maximum $\log d$, if and only if $\Phi$ is a unitary channel.

Thus, the temporal correlation is the strongest, that is, $Q(\Phi)=\log d$ if and only if $\Phi$ is a quantum unitary channel.
Physically, unitary channels describe evolutions of closed quantum systems.  In other words, for a quantum unitary channel, in principle, we can decide the quantum state at an earlier instant from the present one.
The resulting quantity $Q(\cdot)$ fits to common intuition.

\

\textit{The weakest temporal correlations:} If the effect of temporal correlation $\Phi$ is the coherence destroying channel $\Psi$~\cite{zw}, then, for each initial state $\rho=\sum_{i,j}\rho_{ij}E_{ij}$ at $t_{0}$, the resulting state $\Psi(\rho)$ will lose  all of its information about coherence.
That is, $$\Psi(E_{ij})=0,  i\neq j,$$ and $$\Psi(E_{ii})=\sum_{j}p^{i}_{j}E_{jj}, \forall i,$$ where $\{p^{i}_{j}\}_{j}$ is a probability distribution for each $i$.
In this setting, $\rho_{\mu,\Psi}$ is a classical-classical state~\cite{cc1, cc2}. Because the entanglement of formation of classical-classical states are $0$, thus $Q(\Psi)=0$.
In this scenario, the temporal correlation $\Psi$ makes all coherence information lost.
As a result,  it is impossible to decide the  quantum state at an earlier instant from the present one.

\

\textit{The general temporal correlations:} In addition to the strongest and the weakest,  in general, one can have the temporal correlation $\Phi$ in between, with the entanglement formation $Q(\Phi)\in(0, \log d)$.
By expressing $Q(\Phi)$ as
\begin{eqnarray}
Q(\Phi)=\inf_{U}E(\Phi\otimes I(U\otimes U\frac{1}{\sqrt{d}}\sum_{i}\ket{\alpha_{i}\alpha_{i}})),
\end{eqnarray}
with $U$ running over all possible unitary operators on $H_{0}$, we can  use $\sigma_{U}$ to represent the quantum state $U\otimes U\frac{1}{\sqrt{d}}\sum_{i}\ket{\alpha_{i}\alpha_{i}}$.
Now, let $\mathcal{S}$ be the set of all separable states on $H_{1}\otimes H_{0}$ and the norm $\|\cdot \|$ is the trace norm $\|X\|=\text{Tr}(\sqrt{X^{\dagger}X})$.
The distance between a maximally entangled state $\sigma$ and $\mathcal{S}$ can be defined as~\cite{dist1, dist2}
\begin{eqnarray}
l\doteq \text{min}_{\omega\in \mathcal{S}}\|\sigma-\omega\| > 0.
\end{eqnarray}
For each unitary operator $U\in L(H_{0})$, the state $\sigma_{U}$  is always a maximally entangled state.
The corresponding distance between $\sigma_{U}$  and $\mathcal{S}$ is also $l$.
We want to remark that the entanglement formation $E(\cdot)$ is continuous with respect to trace norm~\cite{formation1, formation2, formation3, formation4} and the distance $l$ of a maximally entangled state $\sigma$ and $\mathcal{S}$ is larger than $0$.
As a result,  for the set of maximally entangled states and a constant $c\in (0, \log d)$, one can always find a $\delta >0$  such that if the distance between a state $\sigma$ and a maximally entangled state is smaller than $\delta$, then $E(\sigma)>c$.

Now, let $V\in L(H_0, H_1)$ be a unitary operator, $\mathcal{V}(\rho)=V\rho V^{\dagger}$ be the corresponding unitary channel, and $\varepsilon\in [0,1]$. Consider the quantum channels $\mathcal{V}_{\varepsilon}$:
\begin{equation}
\mathcal{V}_{\varepsilon}(\rho)=(1-\varepsilon)\mathcal{V}(\rho)+\varepsilon\frac{1}{d}I.
\end{equation}
One can see that  if $\varepsilon\neq 0$,  the channel $\mathcal{V}_{\varepsilon}$ is not a unitary channel.

Now, for quantum channels, we introduce the diamond norm, marked as  $\|\cdot\|_{\diamondsuit}$~\cite{diamond1, diamond2}.
That is, $\|\Phi\|_{\diamondsuit}\doteq \max_{\zeta}\|\Phi\otimes I(\zeta)\|$, where $\zeta$ runs over all possible states on $H_{1}\otimes H_{0}$. For quantum channels $\mathcal{V}_{\varepsilon}$ , we have
\begin{eqnarray}
\|\mathcal{V}_{\varepsilon}\otimes I(\sigma_{U})-\mathcal{V}\otimes I(\sigma_{U})\|\leq \|\mathcal{V}_{\varepsilon}-\mathcal{V}\|_{\diamondsuit}.
\end{eqnarray}
When $\varepsilon$ is small enough, we have the following inequality:
\begin{eqnarray}\|\mathcal{V}_{\varepsilon} \otimes I(\sigma_{U})-\mathcal{V} \otimes
I(\sigma_{U})\|\leq \delta, \forall\ \,\, U.
\end{eqnarray}
For all $U$ on $H_{0}$, we already know that $\mathcal{V}\otimes
I(\sigma_{U})$ is a maximally entangled state.
Therefore, for all the unitary operator $U$ on $H_{0}$, we have
$E(\mathcal{V}_{\varepsilon}(\sigma_{U}))> c>0$.
On the other hand, if $\varepsilon\neq 0$, we have known that $\mathcal(V)_{\varepsilon}$ is not a unitary channel, resulting in  $Q(\mathcal(V)_{\varepsilon})< \log d$.
Therefore, by choosing a suitable  $\varepsilon$, one can have $c\leq
Q(\mathcal{V}_{\varepsilon})=\inf_{U}E(\mathcal{V}_{\varepsilon}\otimes
I(\sigma_{U}))<\log d$.

\

In conclusion, based on the entangled-history formalism, we not only affirmatively answer the long-standing question on what quantum temporal correlations are, but also give new insights to the temporal correlations.
Temporal correlations are equivalent to spatial correlations and different instants can be seen as different places.
What is  more,  with the help of quantum state tomography technique,  temporal correlations can be determined completely.
Based on this framework, we also illustrate the strongest and the weakest scenarios, as well as the general ones, by calculating the corresponding entanglement formation.
A uniform aspect on the temporal correlations to our physical intuitions can be naturally obtained, but with a deeper understanding on the link between quantum channels and temporal correlations.

We thank Chung-Yun Hsieh for his important suggestions and remarks for the manuscript. This work is supported by
National Natural Science Foundation of China under Grant No.
11571307, 61877054 and 11675113, Ministry of Science and Technology
of Taiwan (109-2112-M-007-019-MY3), the Key Project of Beijing
Municipal Commission of Education (KZ201810028042), and
Beijing Natural Science Foundation (Z190005). I. Chakrabarty and M.
Shukla acknowledge Zhejiang University for supporting their visits.

\

\end{document}